\documentclass[a4paper,11pt]{article}
\usepackage{pos}

\usepackage{physics}
\usepackage{simplewick}

\newcommand\beq{ \begin{eqnarray} }
\newcommand\eeq{ \end{eqnarray} }

\usepackage[legacycolonsymbols]{mathtools}

\usepackage{color}

\title{New configuration set of HAL QCD collaboration}
\author*[a,b]{Etsuko Itou for HAL QCD collaboration}
\affiliation[a]{Yukawa Institute for Theoretical Physics, Kyoto University, Kitashirakawa Oiwakecho, Sakyo-ku, Kyoto 606-8502, Japan}

\affiliation[b]{Interdisciplinary Theoretical and Mathematical Sciences Program (iTHEMS), RIKEN, Wako, Saitama 351-0198, Japan}

\emailAdd{itou@yukawa.kyoto-u.ac.jp}
\abstract{
We give a brief report on the basic properties and cutoff scale of our new configuration set (HAL-Conf-2023).
We generated 8,000 trajectories of the gauge configurations on $96^4$ lattices with the same lattice parameters as the PACS collaboration~\cite{Ishikawa:2018jee, PACS:2019ofv}.
The topological distribution, the PCAC masses, and decay constants for pseudo-scalar mesons are studied.
As for the scale setting, we utilize the $\Omega$ baryon mass as a reference scale and carefully investigate the operator dependence of the correlation function.
As a result, we obtain $a^{-1}=2338.8(1.5)^{+0.2}_{-3.3} \ [{\rm MeV}]$ as a lattice cutoff. Our hadron spectra in the physical unit reproduce well the experimental results.
}

\FullConference{The 40th International Symposium on Lattice Field Theory (Lattice 2023)\\
July 31st - August 4th, 2023\\
Fermi National Accelerator Laboratory\\}

\begin{document}
\maketitle

\section{Introduction}
This is a brief report on our new configuration set at the physical point by HAL QCD Collaboration.
We employ $N_f$=2+1 nonperturbatively ${\cal O}(a)$-improved Wilson quark action with stout smearing
and the Iwasaki gauge action at $\beta = 1.82$ on a $L^4=96^4$ box.
PACS collaboration recently performed similar physical point generation
with the same action and the same $\beta$ (the same cutoff) but with two different volumes, $L^4= 64^4$ and $128^4$.
The latter of which is called ``PACS10'' configurations~\cite{Ishikawa:2018jee, PACS:2019ofv}.
With their revised determination of the cutoff, $a^{-1} = 2316.2(4.4)$ [MeV],
the simulation point for PACS10 configurations is found to be $(m_\pi, m_K) = (135.3(6)(3), 497.2(2)(9))$ [MeV].

We generated $8,000$ Monte Carlo trajectories and it is very high statistics.
The basic properties of the configuration set, {\it{e.g.}} the PCAC masses, the decay constants of pseudoscalar mesons,
and topological distribution are investigated.
The scale setting of the configuration set is evaluated from the $\Omega$ baryon spectrum though the PACS collaboration utilized the $\Xi$ baryon spectrum.
The result is $a^{-1}=2338.8(1.5)^{+0.2}_{-3.3} \ [{\rm MeV}]$.
To obtain a clear signal of plateau of the effective mass analysis for $\Omega$ baryon, we use the wall source to measure its correlation function and carefully investigate the operator dependence of the plateau value. 

Numerical computations are performed on the supercomputer Fugaku,
the new flagship supercomputer in Japan which succeeds the K computer.
We refer to our new configuration set as ``HAL-Conf-2023''.
We will report the details of the analyses for HAL-Conf-2023 in a full paper near future~\cite{paper-Fconf}.
In this manuscript, we give a brief summary of that.

%%%%%%%%%%%%%%%%%%%%%%%%%%%%%%%%%%%%%%%%%%%%%%%%%%%%%
%%%%%%%%%%%%%%%%%%%%%%%%%%%%%%%%%%%%%%%%%%%%%%%%%%%%%
%%%%%%%%%%%%%%%%%%%%%%%%%%%%%%%%%%%%%%%%%%%%%%%%%%%%%
\section{Simulation setup for configuration generations and basic properties}
%%%%%%%%%%%%%%%%%%%%%%%%%%%%%%%%%%%%%%%%%%%%%%%%%%%%%
%%%%%%%%%%%%%%%%%%%%%%%%%%%%%%%%%%%%%%%%%%%%%%%%%%%%%
%%%%%%%%%%%%%%%%%%%%%%%%%%%%%%%%%%%%%%%%%%%%%%%%%%%%%
We have generated $2+1$ flavor QCD configurations employing the Iwasaki gauge and $\mathcal{O}(a)$-improved Wilson-clover quark actions. 
The lattice extent is $96^4$ and the lattice bare coupling constant $\beta=6/g^2=1.82$ following Refs.~\cite{Ishikawa:2018jee, PACS:2019ofv}.
The gauge field is $6$ times smeared using the stout smearing parameter $\rho=0.1$.
We utilize $c_{SW}=1.11$, which is nonperturbatively determined by the Schr\"{o}dinger functional scheme in Ref.~\cite{Taniguchi:2012gew}.
The hopping parameters for u,d quarks and s quark are set to $(\kappa_{ud},\kappa_s)=(0.126117,0.124902)$, then the hadron masses are then reported to be almost the values at the physical point in Refs.~\cite{Ishikawa:2018jee, PACS:2019ofv}.

To reach the physical point, we perform $5$-run series in the configuration generation process, and in each run, the generation was started with different random number seeds and different old configurations, in which we took heavier s-quark mass parameter; $\kappa_s=0.12479$~\cite{Ishikawa:2015rho}.
After discarding more than $300$ trj. to remove the thermalization process in each run series, we generate $1,600$ trj. in each run, thus, $8,000$ trj. in total. We save the configurations every $5$ trajectory and use them for the measurement of physical observables, {\it e.g.} topological charge, and hadron correlation functions.

%%%%%%%%%%%%%%%%%%%%%%%%%%%%%%%%%%%%%%%%%%%%%%%%%%%%%
%%%%%%%%%%%%%%%%%%%%%%%%%%%%%%%%%%%%%%%%%%%%%%%%%%%%%
\subsection{Plaquette value and topological charge}
\begin{figure}[h]
    \centering
	    \includegraphics[width=0.8\textwidth]{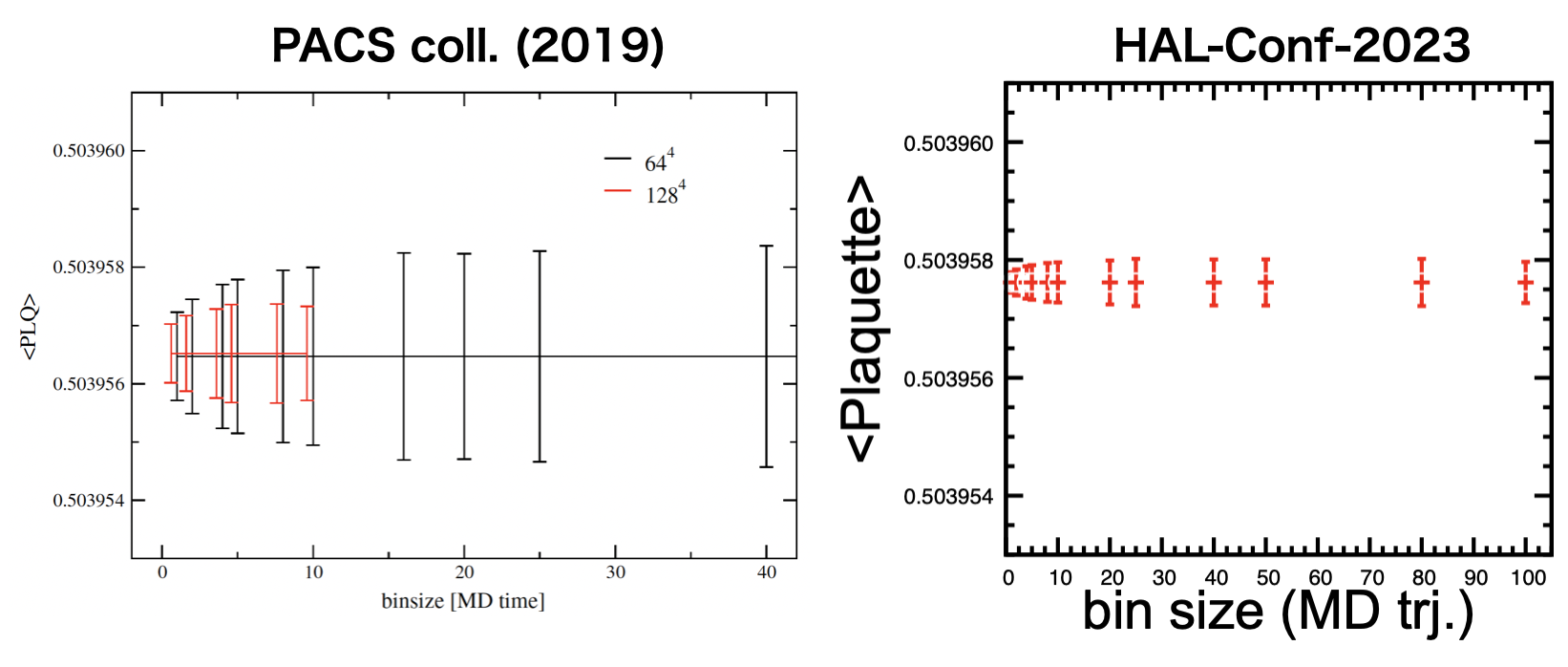}
	    \caption{Bin-size dependence in the jackknife analysis for the plaquette average (Left) Result by PACS collaboration at same lattice parameters, $\beta, \kappa_{ud}$, and $\kappa_s$, on $64^4$ and $128^4$ lattices. The figure is originally given in Ref.~\cite{Ishikawa:2018jee}. (Right) Our result on $96^4$ lattice.}
	    \label{fig:comp-plaq-PACS}
\end{figure}

%%%%%%%%%%%%%%%%%%%%%%%%%%%%%%%%%%%%%%%%%%%%%%%%%%%%%
%%%%%%%%%%%%%%%%%%%%%%%%%%%%%%%%%%%%%%%%%%%%%%%%%%%%%
First, let us show the averaged plaquette value and its bin-size dependence on the jackknife error.
Figure~\ref{fig:comp-plaq-PACS} is the comparison plots between the data by PACS collaboration on $64^4$ and $128^4$ lattices with the same $\beta$ and $\kappa_{ud},\kappa_s$~\cite{Ishikawa:2018jee, PACS:2019ofv} and ours on $96^4$ lattice.  
We take the bin-size$=100$ trajectories and then obtain $\langle \mathrm{plaquette} \rangle = 0.5039576(3)$.
Our results are $1$-$\sigma$ consistent with the data from PACS collaboration but the statistical errors are now very small.

The size of statistical error is strongly related to the auto-correlation time, so that we next investigate it. Although the auto-correlation time depends on the measurement quantity, in general, the physical quantity related to the low mode has a long auto-correlation time. Here, we show one of such quantities, namely the topological charge. 
%%%%%%%%%%%%%%%%%%%%%%%%%%%
%%%%%%%%%%%%%%%%%%%%%%%%%%%

To measure the topological charge, we adopt the gluonic definition through the gradient flow~\cite{Luscher:2010iy}.
In our work, we utilize the standard plaquette gauge action as a flow gauge action, and to solve the flow equation the third-order Runge-Kutta algorithm with the step size $\epsilon =0.01$ is adapted.
The value of $Q(t)$ roughly plateaus at a long flow-time ($t$) regime, but small fluctuations exist. Therefore, we introduce a reference scale $t_0$ and identify the value of $Q(t=t_0)$ as a convergent value of $Q$ for each configuration~\cite{Bruno:2014ova}.
Here, the reference scale $t_0$ is originally introduced in Ref.~\cite{Luscher:2010iy} defined as $t^2 \langle E(t) \rangle |_{t=t_0} =0.3$.
The result of the reference scale in our simulation is  $t_0/a^2 = 2.1047(4)$.

\begin{figure}[h]
    \centering
	    \includegraphics[width=0.6\textwidth]{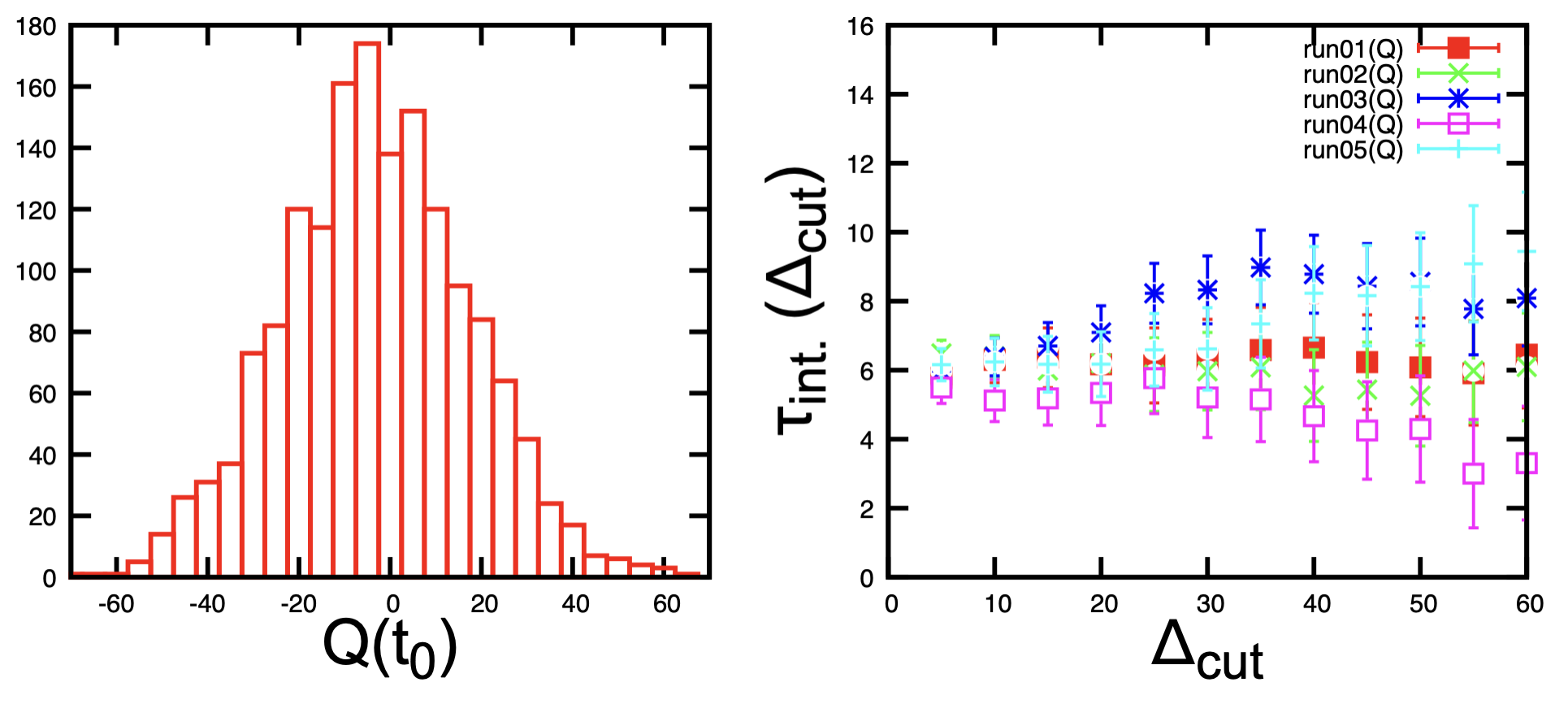}
	    \caption{(Left) Topological charge distributions for our HAL-Conf-2023 (Right) The integrated auto-correlation time for the topological charge as a function of the cutoff MD trj.( $\Delta_{\mathrm{cut}}$ ).}
	    \label{fig:Q-and-AC}
\end{figure}

The left panel of Figure~\ref{fig:Q-and-AC} represents the histogram of $Q(t_0)$ for a total $1,600$ generated configurations.
%%%%%%%%%%%%%%%%%%%%%%%%%%%
%%%%%%%%%%%%%%%%%%%%%%%%%%%
It shows almost a Gaussian distribution as expected.
We obtain the averaged value $\langle Q \rangle=-0.6644(6167)$, $\langle Q^2 \rangle=416.8(16.1)$, and the susceptibility $\chi_Q= \langle (Q-\langle Q \rangle)^2/V=4.90 (19)  \times 10^{-6}$.
As we expected, $\langle Q \rangle $ is consistent with zero within $1.1$--$\sigma$.
The right panel of Figure~\ref{fig:Q-and-AC} displays the integrated auto-correlation time for $Q(t_0)$. Here, we follow the error estimation method proposed by RBC/UKQCD collaboration~\cite{RBC:2012cbl, RBC:2014ntl}.
We analyze our data for each run series independently. The obtained integrated auto-correlation time saturates in a short trajectory regime, namely around $12$ MD trajectories. 

Although such a short auto-correlation may seem unusual, in fact, the value of $Q$ changes significantly even for $1$ configuration separation.
There are two possible reasons for this short auto-correlation time.
One is due to the usage of Wilson fermion, where the near-zero mode is absent even in a small quark mass regime. The other is due to the relatively coarse lattice spacing of $a\approx 0.08$[fm] in our simulations, which will be obtained later.
As a consequence, the short auto-correlation time suggests that our set of $1,600$ configurations is well spread in the configuration space, which is a rather good set.

%%%%%%%%%%%%%%%%%%%%%%%%%%%%%%%%%%%%%%%%%%%%%%%%%%%%%
%%%%%%%%%%%%%%%%%%%%%%%%%%%%%%%%%%%%%%%%%%%%%%%%%%%%%
%%%%%%%%%%%%%%%%%%%%%%%%%%%%%%%%%%%%%%%%%%%%%%%%%%%%%
\section{Hadron correlation functions}
%%%%%%%%%%%%%%%%%%%%%%%%%%%%%%%%%%%%%%%%%%%%%%%%%%%%%
%%%%%%%%%%%%%%%%%%%%%%%%%%%%%%%%%%%%%%%%%%%%%%%%%%%%%
%%%%%%%%%%%%%%%%%%%%%%%%%%%%%%%%%%%%%%%%%%%%%%%%%%%%%
\subsection{Mesonic quantities}
We turn to the mesonic quantities, {\it{e.g.}} the meson spectra, the decay constants of pseudo-scalar (PS) meson, and the PCAC masses for ud- and s-quark.
To measure the correlation function for the PS and axial-vector current ($A_\mu$) operators, we employ the wall source method without gauge fixing~\cite{Kuramashi:1993ka}.
To suppress statistical fluctuations, we take the average between forward and backward propagation and we also measure the correlation function in each of the four directions for one configuration using the fact that our lattice is hypercubic.
The statistical errors in this section are also estimated by the jackknife method.

By the simultaneous fit of the PS-PS and PS-A$_4$ correlation functions for the light-light quark and the light-heavy quark sectors, we can obtain the PS meson masses, the decay constants, and the PCAC mass for each quark. We will show the results of PS meson masses later.
Here, we show the other quantities.

\begin{figure}[h]
    \centering
	    \includegraphics[width=0.7\textwidth]{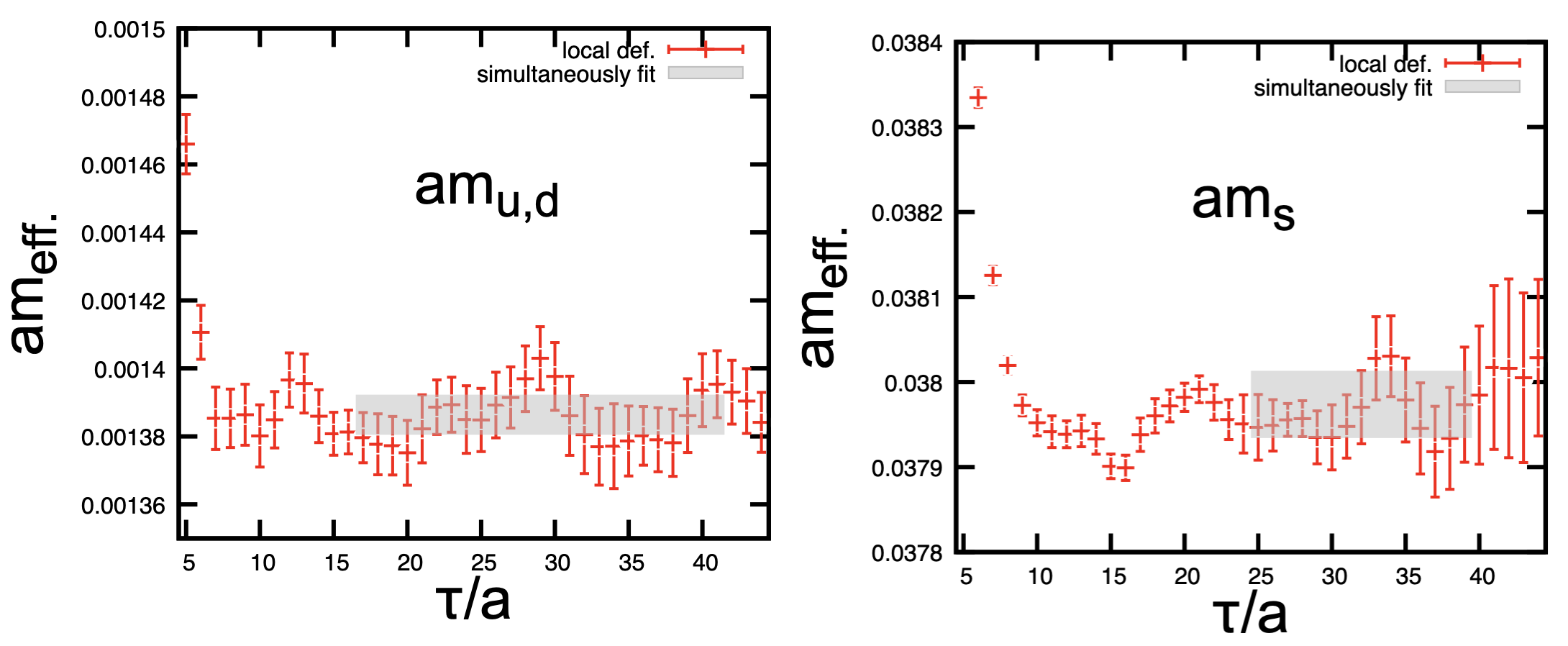}
	    \caption{PCAC masses for u,d-quark (left panel) and s-quark (right panel). }
	    \label{fig:PCAC-masses}
\end{figure}

As for the PCAC mass, we also investigate the effective masses, which are obtained by taking the ratio of 2-data points of the correlation function, as a function of $\tau/a$.
Figure~\ref{fig:PCAC-masses} shows the PCAC masses for ud- and s-quark in lattice unit.
%%%%%%%%%%%%%%%%%%%%%%%%%%%
%%%%%%%%%%%%%%%%%%%%%%%%%%%
Here, we also plot the results and their fitting range in the simultaneous-fit analyses as the shadowed regimes.
We can see that the effective mass fluctuates around the shadowed regime and the width of the fluctuation also matches the one of the shadow reasonably. Therefore, we conclude that the simultaneous-fit analyses and the effective mass analyses are consistent with each other.

\begin{figure}[h]
    \centering
	    \includegraphics[width=0.45\textwidth]{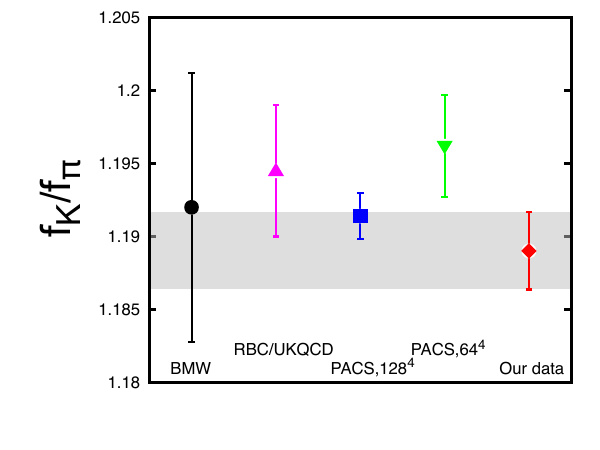}
	    \caption{The ratio of decay constants $f_K/f_{\pi}$.}
	    \label{fig:fK-fpi}
\end{figure}

Figure~\ref{fig:fK-fpi} is a summary plot of the ratio of decay constants, $f_K/f_\pi$ for $2+1$-flavor QCD.
%%%%%%%%%%%%%%%%%%%%%%%%%%%
%%%%%%%%%%%%%%%%%%%%%%%%%%%
Here, we compare our result with the data given by BMW~\cite{Durr:2010hr}, RBC/UKQCD~\cite{RBC:2014ntl}, PACS ($128^4$), PACS ($64^4$ with reweighted)~\cite{PACS:2019ofv}. It seems that our data is almost consistent with these results.

%%%%%%%%%%%%%%%%%%%%%%%%%%%%%%%%%%%%%%%%%%%%%%%%%%%%%
%%%%%%%%%%%%%%%%%%%%%%%%%%%%%%%%%%%%%%%%%%%%%%%%%%%%%
\subsection{Calculation strategy of baryonic quantities}
%%%%%%%%%%%%%%%%%%%%%%%%%%%%%%%%%%%%%%%%%%%%%%%%%%%%%
%%%%%%%%%%%%%%%%%%%%%%%%%%%%%%%%%%%%%%%%%%%%%%%%%%%%%
The effective mass of a baryon is evaluated from the correlators with point-sink and wall-source.
One of the reasons for using the wall source correlator is that
the meson-baryon scattering state is expected to be suppressed 
in wall source state because the lowest meson-baryon state is the p-wave state. 
In addition, 
in the context of inter-baryon interactions, 
the wall source is found to have a large overlap with the ground state 
\cite{Iritani:2019}.

Furthermore, for the consistency check, we also utilize the variational method with smeared source and smeared sink operators.  To resolve the mixing between the ground and excited states, we solve a generalized eigenvalue problem (GEVP) for the correlator matrix of baryons with a suitable smearing function.
The smearing function in this work is taken as the Gaussian-type function
\beq
f(r)=f_0{\rm e}^{-\alpha r^2}.
\eeq
We choose the smearing parameters $\alpha$ for the narrow and broad extents of the quark fields, 
corresponding to the root mean square of the radius 
$\sqrt{\langle r^2 \rangle}=0.20 {\rm fm}$ and 
$\sqrt{\langle r^2 \rangle}=0.68 {\rm fm}$, respectively.

\subsection{Scale setting}
We set the cutoff scale from the $\Omega$ baryon mass, while the PACS collaboration utilizes the $\Xi$ baryon mass.
As a reference scale, we utilize the experimental data of the $\Omega$ baryon mass
from Particle Data Group, 
\beq
m_{\Omega}=1672.45 \ [{\rm MeV}].\label{eq:PDG-data}
\eeq

\begin{figure}[htbp]
 \begin{center}
        \includegraphics[keepaspectratio, scale=0.3]{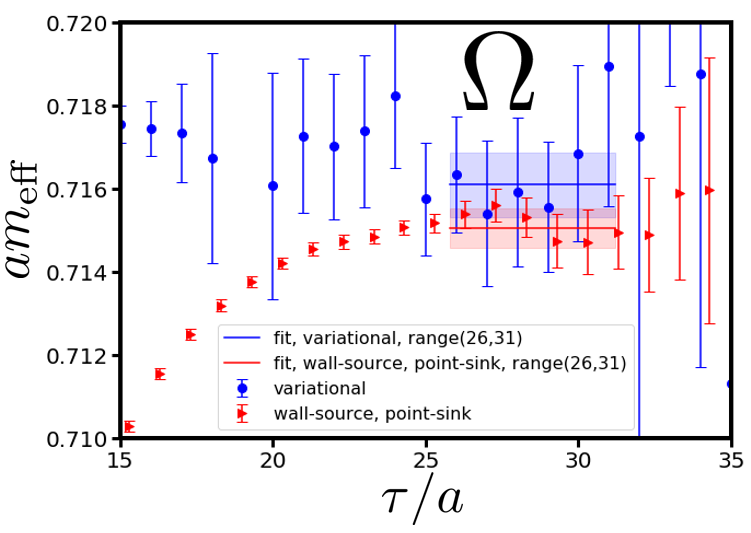}
    \caption{Effective mass for $\Omega$ baryon in the lattice unit with two methods: The wall-source and point-sink operators (triangle-red symbols) and the variational method (circle-blue symbols).
}\label{fig:effective-mass-Omega}
\end{center}
  \end{figure}

Our results of the effective mass of $\Omega$ baryon from the wall-source and variational method 
are shown in Figure~\ref{fig:effective-mass-Omega}. 
The plateau values expressed as a shadowed regime of the two methods are consistent. 
We adopt the results from wall-source data as a central results and to estimate the systematic error we utilize the results from the variational method. 
%%%%%%%%%%%%%%%%%%%%%%%%%%%%%%%
 %%%%%%%%%%%%%%%%%%%%%%%%%%%%%%%%%%%%%
Compared with the reference value~\eqref{eq:PDG-data}, we fix the physical scale of our lattice cutoff as
\beq
a&=0.084372(54)^{+118}_{-6} \ [{\rm fm}], \quad a^{-1}&=2338.8(1.5)^{+0.2}_{-3.3} \ [{\rm MeV}].
\eeq

\subsection{Hadron spectra}
\begin{figure}[h]
    \centering
	    \includegraphics[width=0.45\textwidth]{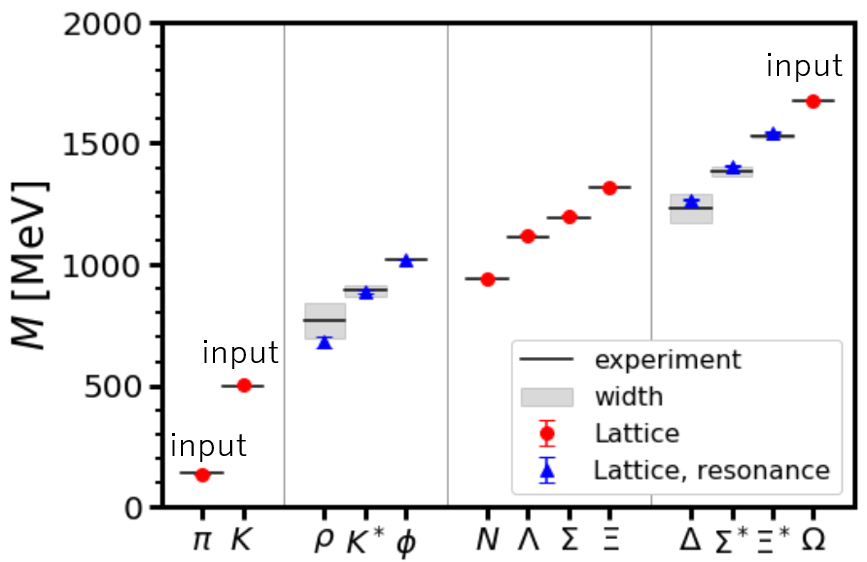}
	    \caption{The summary plot of hadron spectra in physical unit.}
	    \label{fig:summaryfig}
\end{figure}

Finally, Figure~\ref{fig:summaryfig} depicts the summary plot of hadron spectra in physical unit.
Here, we also show the masses of the low-lying hadrons including the resonance states, 
$\rho, K^*, \Delta, \Sigma^*, \Xi^*$, 
which are estimated from the naive effective mass analysis using the wall-source data. 
%%%%%%%%%%%%%%%%%%%%%%%%%%%
%%%%%%%%%%%%%%%%%%%%%%%%%%%

%%%%%%%%%%%%%%%%%%%%%%%%%%%%%%%%%%%%%%%%%%%%%%%%%%%%%
%%%%%%%%%%%%%%%%%%%%%%%%%%%%%%%%%%%%%%%%%%%%%%%%%%%%%
%%%%%%%%%%%%%%%%%%%%%%%%%%%%%%%%%%%%%%%%%%%%%%%%%%%%%
\section{Summary}
%%%%%%%%%%%%%%%%%%%%%%%%%%%%%%%%%%%%%%%%%%%%%%%%%%%%%
%%%%%%%%%%%%%%%%%%%%%%%%%%%%%%%%%%%%%%%%%%%%%%%%%%%%%
In this proceeding, we give a brief report on the basic properties of the new configuration set (HAL-Conf-2023) generated by HAL QCD collaboration.
We employ the same lattice setup with the PACS collaboration~\cite{Ishikawa:2018jee, PACS:2019ofv}, but here we generated 8,000 trj. on $96^4$ lattices.
As for the scale setting, we utilize the $\Omega$ baryon mass as a reference scale. Here, we take the wall-point source correlation function to improve the signal of the effective mass and carefully investigate the operator dependence.
As a result, we obtain $a^{-1}=2338.8(1.5)^{+0.2}_{-3.3} \ [{\rm MeV}]$ and our hadron spectra in the physical unit reproduce well the experimental results.
The detailed report will appear in Ref.~\cite{paper-Fconf}.

%%%%%%%%%%%%%%%%%%%%%%%%%%%%%%%%%%%%%%%%%%%%%%%%%%%%%

%%%%%%%%%%%%%%%%%%%%%%
\acknowledgments
We thank I.~Kanamori, K.-I.~Ishikawa, N.~Ukita, and members of the HAL QCD Collaboration for stimulating discussions.
The work of E.~I. is supported by JSPS KAKENHI with Grant Number 23H05439,%Kiban-S
JST PRESTO Grant Number JPMJPR2113,%Sakigake
JSPS Grant-in-Aid for Transformative Research Areas (A) JP21H05190, %ExU
JST Grant Number JPMJPF2221  % SQAI
and also supported by Program for Promoting Researches on the Supercomputer ``Fugaku'' (Simulation for basic science: from fundamental laws of particles to creation of nuclei) and (Simulation for basic science: approaching the new quantum era), and Joint Institute for Computational Fundamental Science (JICFuS), Grant Number JPMXP1020230411.%Fugaku
This work is supported by Center for Gravitational Physics and Quantum Information (CGPQI) at YITP. 

%%%%%%%%%%%%%%%%%%%%%%%
%%%%%%%%%%%%%%%%%%%%%%%
%%%%%%%%%%%%%%%%%%%%%%%
%\bibliographystyle{utphys}
\bibliographystyle{utphys_short}
\bibliography{./latticeQCD.bib}
%%%%%%%%%%%%%%%%%%%%%%%
%%%%%%%%%%%%%%%%%%%%%%%
%%%%%%%%%%%%%%%%%%%%%%%

%\begin{thebibliography}{99}
%\bibitem{...}
%....

%\end{thebibliography}

\end{document}